\documentclass[conference]{IEEEtran}

\IEEEoverridecommandlockouts
\usepackage{cite}
\usepackage{amsmath,amssymb,amsfonts}
\usepackage{graphicx}
\usepackage{textcomp}
\usepackage{xcolor}
\usepackage[font=small,labelfont=bf]{caption}
\usepackage{fancyhdr}
\usepackage{hyperref}
\newcommand{\ignore}[1]{}

\usepackage{adjustbox}
\usepackage{amsmath} 
\usepackage{amssymb}
\usepackage{subfig}
\usepackage{etoolbox}
\usepackage{graphicx}
\usepackage{float}
\usepackage{multirow}
\usepackage{hhline}
\usepackage[english]{babel}
\usepackage[autostyle]{csquotes}
\usepackage[utf8]{inputenc}
\usepackage{dirtytalk}
\usepackage[noend]{algpseudocode}
\graphicspath{ {images/} }

\def\BibTeX{{\rm B\kern-.05em{\sc i\kern-.025em b}\kern-.08em
    T\kern-.1667em\lower.7ex\hbox{E}\kern-.125emX}}

\title{\LARGE \bf DeCoILFNet: Depth Concatenation and Inter-Layer Fusion based ConvNet Accelerator}

\author{Akanksha Baranwal $^{1,*}$, Ishan Bansal$^{1,*}$, Roopal Nahar$^{1}$, K.Madhava Krishna$^{1}$

\thanks{$^{1}$ Robotics Research Center, IIIT-Hyderabad, India}%
\thanks{$^{*}$ Equal contribution}%
\thanks{{\tt\small akankshabar@gmail.com}}%
\thanks{{\tt\small bansalishan0@gmail.com}}%
\thanks{{\tt\small roopalnahar08@gmail.com}}%
\thanks{{\tt\small mkrishna@iiit.ac.in}}%
}

\begin{document}

\maketitle
\thispagestyle{empty}
\pagestyle{plain}

\begin{abstract}
Convolutional Neural Networks (CNNs) are rapidly gaining popularity in varied fields. Due to their increasingly deep and computationally heavy structures, it is difficult to deploy them on energy constrained mobile applications. Hardware accelerators such as FPGAs have come up as an attractive alternative. However, with the limited on-chip memory and computation resources of FPGA, meeting the high memory throughput requirement and exploiting the parallelism of CNNs is a major challenge. We propose a high-performance FPGA based architecture - \textit{Depth Concatenation and Inter-Layer Fusion based ConvNet Accelerator - DeCoILFNet} which exploits the intra-layer parallelism of CNNs by flattening across depth and combines it with a highly pipelined data flow across the layers enabling inter-layer fusion. This architecture significantly reduces off-chip memory accesses and maximizes the throughput. Compared to a 3.5GHz hexa-core Intel Xeon E7 caffe-implementation, our 120MHz FPGA accelerator is 30X faster. In addition, our design reduces external memory access by 11.5X along with a speedup of more than 2X in the number of clock cycles compared to state-of-the-art FPGA accelerators.
\end{abstract}

\section{Introduction}
From recognition to reasoning, convolution neural networks have attained impressive accuracies in a broad range of applications such as mobile robotics, natural language processing, information retrieval and speech recognition. \cite{fcn} \cite{segnet}. In 2014, VGG-Net, \cite{VGGNet} a network which became very popular suggested some standards including uniform filters/kernels of size 3X3 across all layers as it could emulate the effect of larger receptive fields. This reinforced the notion that convolution neural networks have to be deep in order for the hierarchical representation of visual data to work. 
\par General purpose processors are not able to fully exploit the inherent inter-output and intra-output parallelism of convnet networks, hence specialized hardware accelerators such as GPUs\cite{gpu}, FPGA\cite{dynamically configurable coprocessor} and ASICs\cite{asic} are gaining popularity. In fields like mobile robotics, which usually have stringent  energy constraints, the reconfigurability and higher energy efficiency of FPGA based implementations has made them an attractive alternative. \cite{Fused layer cnn} \cite{optimizing fpga}
  \begin{figure}[h]
\centering
\minipage{0.45\textwidth}
  \includegraphics[width= \linewidth,height = 3 cm]{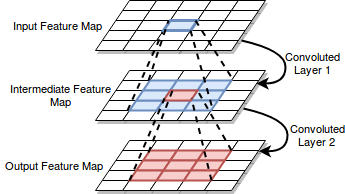}
  \endminipage\hfill
\caption{Data influence diagram across layers: \emph{For computing subsequent layers, each element of the input is needed only for a small region of output.}}\label{fig:teaser}
\end{figure}
 The major bottleneck while implementing huge networks on FPGA is meeting high memory throughput requirement of CNNs with limited on-chip memory. Traditional implementations of CNNs evaluate the network layer by layer\cite{dynamically configurable coprocessor} and off-load data intermittently to a larger external memory which significantly decreases throughput because of limited data transfer bandwidth. 
 \par The computation pattern of CNNs is similar to iterative stencil loops (ISLs) \cite{isl}, for which data dependencies span across multiple layers and iterations. Convnet layers are characterised by uniform spatial dependencies, domain narrowness and uniform inter-iteration dependencies. Works like \cite{Fused layer cnn} have adapted ISLs computation techniques \cite{isl} to pipeline the dataflow across different convnet layers. Since the spatial data flow across layers is dependent on very few data values, it is not required to wait for the entire intermediate output to be computed to start processing the next layer. This fact was exploited in Fused layer cnn \cite{Fused layer cnn} which restructured the computation to significantly reduce external memory access. 
\par In our paper, we leverage upon the fact that the reverse is also true. That is, a particular input influences only a limited neighborhood of the intermediate output layers. So once these outputs are computed, that particular input can be discarded as shown in Fig(\ref{fig:teaser}). Using techniques like line buffer windowing and depth based concatenation, our 2.78X faster architecture improves upon \cite{Fused layer cnn}. Specifically we make the following contributions:
\begin{itemize}
\item We propose depth concatenation in both input data and filter weights, i.e. data values across depth are concatenated adjacent to each other so that they can be moved together across buffers.  Since most of the computations along depth for each layer are independent and can occur concurrently, depth flattening minimises the lag due to serial data flow along depth.
\item We have modified the data flow pattern of CNNs for a constrained bandwidth setup by fusing across layers using the architectural pattern of line buffering. Line buffers help maximize data re-use by storing input serial data stream and intermediate computation results in small on-chip BRAM buffers. The effectively pipelined structure allows the computation of values of next layer as  soon as its depending values have been computed and discards this input as soon as the corresponding outputs have been computed thus eliminating recomputation and optimizing memory resources.
\end{itemize}
These contributions have enabled the design of our elegantly pipelined high throughput DeCoILFNet accelerator which is very efficient in terms of utilization of FPGA resources. We have evaluated our accelerator on VGG-like networks, with VGG-16\cite{VGGNet} as the representative. Compared to state-of-the-art CNN FPGA accelerator  \cite{optimizing fpga}, our accelerator performs 2.6X faster on an average and reduces external memory access by 11.5X. Compared to Fused CNN\cite{Fused layer cnn}, our accelerator performs 2.78X faster with a slight increase in  off-chip memory access. We are 30X better in speed compared to CPU-caffe implementation and almost reach the speed of GPU-caffe implementations. 

\section{Related work and Motivation }

There are two major components of computation in Convolutional Neural Networks: forward pass and backward pass. While training iteratively, network performs repeated forward and backward passes to refine weights until the desired accuracy is achieved. Since for recognition only forward pass is required, many application designers train networks offline and use the trained weights to perform time-sensitive jobs on energy constrained devices\cite{Fused layer cnn} \cite{optimizing fpga}. Recent developments in the deep learning community have shown that the fully connected layers can be removed with no degradation in performance \cite{Resnet}. Under these circumstances, works like \cite{optimizing fpga} \cite{dynamically configurable coprocessor} which focus on accelerating convolution layers have gained prominence.
However, as the networks are getting heavier, the on-chip memory of FPGAs is becoming insufficient to store the huge intermediate outputs. 
Conventional works \cite{dynamically configurable coprocessor} have focused on designing CNN accelerators which iteratively process the CNN layers and off-load the intermediate data to external memory. This involves extensive and unnecessary repetitious read and write accesses. Because of this reason, the limited amount of external bandwidth is a challenge for designing efficient accelerators. In CNNs, the input and output feature volume is larger for initial layers and it gradually reduces. In the later layers, the memory occupied by weights dominates as the depth increases. Thus redesigning data flow movement for initial layers significantly reduces the overall external memory accesses\cite{Fused layer cnn} which decreases the overall computation latency and power. 
Inspired by the structure of image processing pipelines to minimize memory bandwidth using architectural pattern of line buffering \cite{darkroom}, our DeCoILFNet uses small on-chip buffers to pipeline the computations within and across the layers increasing throughput and eliminating unnecessary communication with off-chip DDR. Our architecture has been optimized in a bandwidth constrained setup so efficiently that the restricted external memory access is no longer the bottleneck.  

\section { DeCoILFNet Architecture}
In the following sections we describe in detail the optimizations in different modules of  DeCoILFNet accelerator. \

Though our accelerator is generic, for ease of explanation, we have taken the following \textbf{ test example} - \textit{ input image as 5*5*3 (l*b*d), two convolution layers fused both with stride=1 (s), padding=1 (p), number of filters=3 (k) with kernel size 3X3 (wXw) followed by a pooling layer with window of 2X2 and stride =2. }

 \begin{figure}[!tbh]
\centering
\minipage{0.45\textwidth}
  \includegraphics[width=\linewidth,height = 7 cm]{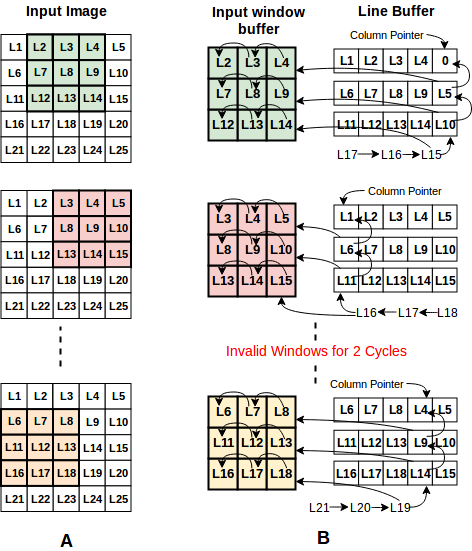}
  \endminipage\hfill
\caption{A: expected input window, B: window obtained from line buffer} \label{fig:linebuffer1}
\end{figure}

\subsection{Line Buffer Windowing Module}
The input to the accelerator comes in the form of a serial data stream. The first layer in CNNs is the convolution layer.  For convolution operation, we need input windows similar to shown in the expected window in Fig. (\ref{fig:linebuffer1}). As the input data comes serially, to get a valid complete window, we need 9 values of the sliding window which come sequentially. This cumulative delay of reading these values for getting a valid window each time adds huge unnecessary delay to overall computation. Therefore our line and window buffer module as shown in Fig. (\ref{fig:linebuffer1}) is pipelined in such a way that we are able to get a new window at each clock cycle after a certain latency.

\begin{figure}[tb]
\centering

  \includegraphics[width=\linewidth,height = 9 cm]{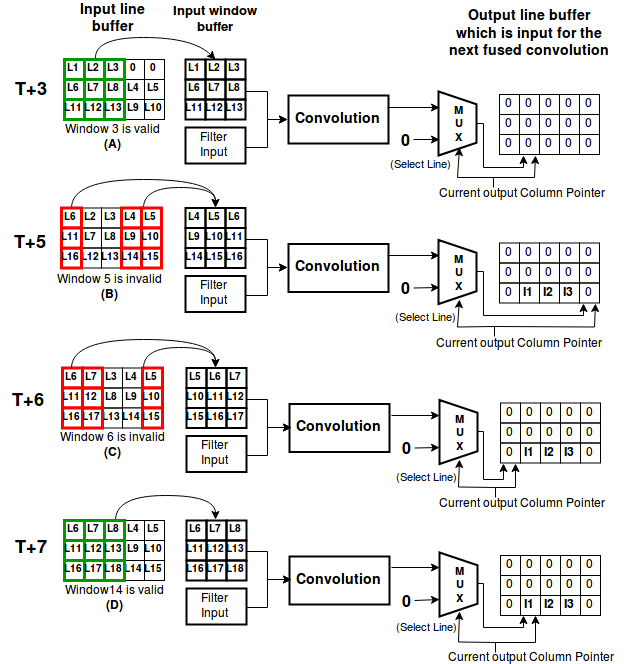}  
\caption{Incorporating padding layer in our architecture using line buffers}\label{fig:padding}
\end{figure}


\par  Usually before convolution, to maintain the spatial dimensions of the output, we pad the input layer with zeros. As shown in Fig. (\ref{fig:padding}), when we reach towards the end of line buffer, we get some invalid windows. Using our line buffer module, we are able to smoothly incorporate the padding layer to get padded windows which are input for the next consecutive convolution.  

\subsection{Depth concatenation module: Input data and filter data flattening}
The above line buffer windowing module has been described for a 2-D window, whereas in our case, for volume convolution, we need a 3-D window. To get the 3-D window similarly in every cycle, our novel method for the same is to flatten along depth so that the data flow is same as before but instead of just one window of a particular depth, we get a window flattened along the third dimension. As shown in the Fig. (\ref{fig:depth_concatenation}), the input data after preprocessed depth-flattening, is sent to DeCoILFNet as a concatenated data stream. This concatenation increases the bandwidth as now instead of reading the 32 bits of D$_{11}$, D$_{21}$, and D$_{31}$ in separate cycles, we read them together as 96 bits of D$_{11}$D$_{12}$D$_{31}$. This new concatenated window can be simply split into three independent windows which are  parallely sent to the convolution block. Data of the convolving filter too is flattened similarly i.e. the values along the depth are concatenated. Before computation, this concatenated data f$_{13}$ of filter1 is split into three 2-D filters and sent to convolution module. 
We have instantiated d*d = 9 filter BRAMs with multiple filters kept one after the other as shown in Fig (\ref{fig:depth_concatenation}). The multiple BRAMs allow us to read all 9 values of one 3-D filter parallely, thus making the filter ready for convolution in one cycle. 

\subsection{3-D Convolution pipelined Module }
\begin{figure}[!tbh]
\centering
\minipage{0.45\textwidth}
  \includegraphics[width=\linewidth,height = 4.5 cm]{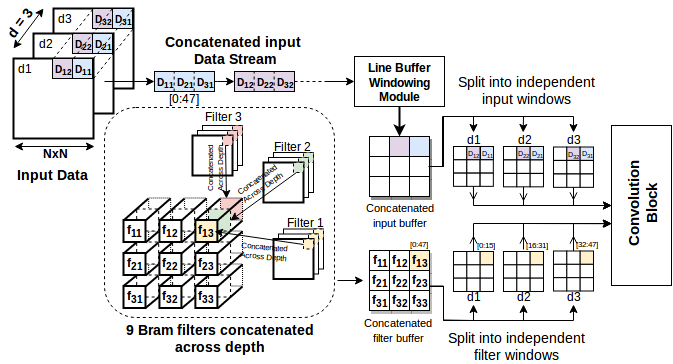}
  \endminipage\hfill
\caption{Depth Concatenation Module for input data and filter} \label{fig:depth_concatenation}
\end{figure}

As shown in Fig. (\ref{fig:depth_concatenation}), the 3-D filter and input window split into d=3 filters and d=3 windows. 
We have used DSPs only for multipliers and LUTs for adders so that more computations can be performed in parallel. Both the multiplier and adder modules have an initial latency of 9 cycles after which because of its internal pipelining, the output of the next k=3  subsequent  filters and input windows keeps coming in every cycle. Thus, the 2-D convolution module gets finely pipelined giving output in every cycle after a latency of (9*(1+ ceil(2(log2)w))) = 45 cycles because of the cumulative effect of multipliers and adders. The d values of 2D-convolution of each filter are added again to give the final single scalar value of 3-D convolution of the output volume. The entire 3-D convolution module is pipelined in such a way that after an initial latency of (9*(1+ ceil(2(log2)w)+ceil(log2(d)))) = 63 cycles we get the output of convolution of each filter with an image window in every clock cycle. \par Activation functions consume a very small percentage in the overall computation and can be trivially integrated without any effect on data flow movement. The ReLU layer which has also been incorporated (without any computation overhead) in this module has not been explicitly shown. 

\subsection{Pooling}
Usually in CNNs, the consecutive convolution layers are followed by pooling. In max pooling, a 2X2 window is slided across the input with a stride of 2. In our DeCoILFNet architecture, we use an intermediate pool line buffer for pipelining. As soon as we get the output of previous convolution, we redirect it to the pool buffer at the current output column address. We update the output column address at every even step, and at odd steps, we replace the current output with the max of old value and new computed output. 
These pooled outputs are read into the next input line buffer for further computation. 

\subsection{Inter-layer Fusion Pipelining}

\begin{figure*}[!tbh]
\centering 
\includegraphics[width=\textwidth,height = 5 cm]{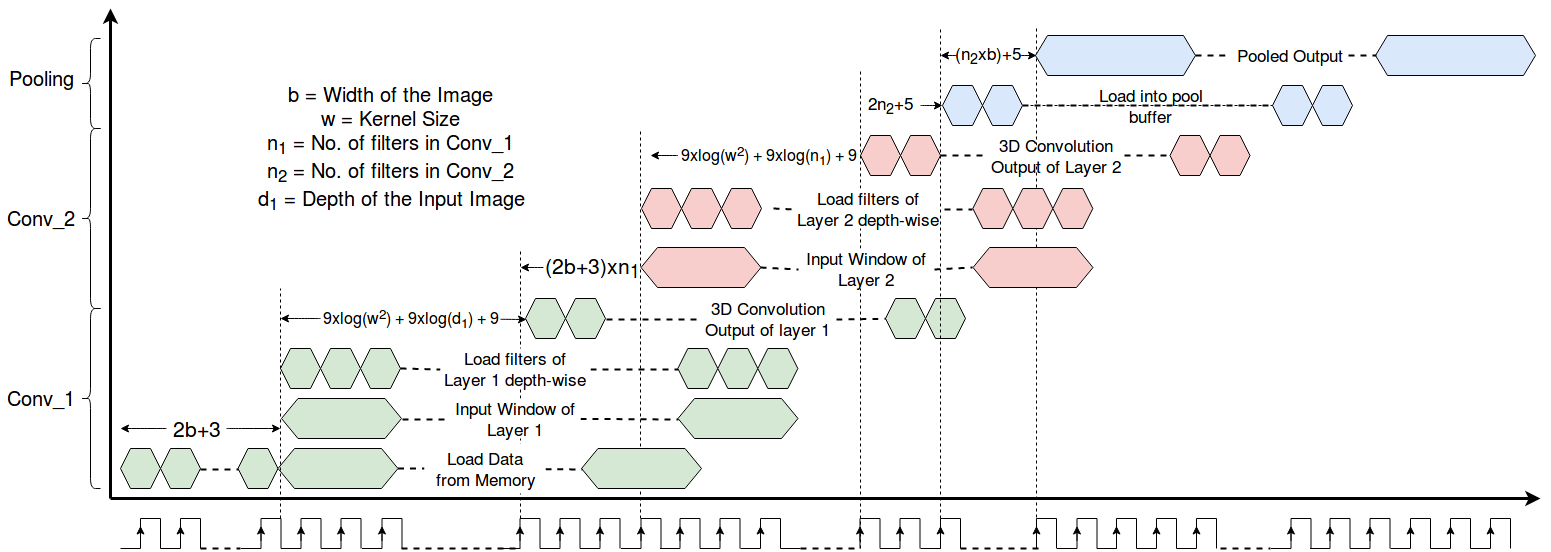}
\caption{Overall Pipeline design}
\label{fig:pipeline}  
\end{figure*}
Since CNNs follow the pattern of iterative stencil loops \cite{isl}, i.e. each  particular input influences only a limited neighborhood of the intermediate output layers as shown in Fig. (\ref{fig:teaser}). The main concept of using line buffer windowing module is based on this idea. So once these outputs are computed, that particular input can be replaced to get the next input either from external memory or computed output of the previous layer. Hence in our architecture, we start processing for the next layer as soon as we get the required valid inputs. As explained above in the 3-D convolution pipelined module, we get the convolution output of intermediate layer in every cycle for filters subsequently one after another. As shown in the pipeline Fig. (\ref{fig:pipeline}), since in the first layer we have three filters which are computed one after another, though we get the output of each filter in every cycle, to stream the output data as serial input to the intermediate layer , we need to wait for the whole volume of output value to be computed. During this time when the volume is being computed, the input window is kept constant till all filters have been processed. This output volume is serially streamed to the intermediate line buffer. Here also, we need to wait for initial filing of intermediate line buffer before we get a valid convolvable window. This pipelining can be continued for further convolution layers. DeCoILFNet accelerator has been pipelined so efficiently that even if multiple convolutions are fused together, the only delay is because of the initial latencies after which we are still able to get one output element in every step. If we fuse the pooling layer in our architecture, as explained above, we need to wait for some more clock cycles before every new pooled row. Hence our architecture works best when we have multiple consecutive convolutions. 
 
\section{Experimental Evaluation and Results}
\subsection{Programming using hardware descriptive language: verilog
} Most of the design optimization works \cite{optimizing fpga} \cite{Fused layer cnn} \cite{dynamically configurable coprocessor} have been done using high level synthesis tools as it is easier to port the code from software to hardware implementation. The motivation behind using HLS is to avoid the need for RTL programming, nevertheless it is still necessary to verify the HLS generated RTL output \cite{HLS31}, and in cases verification fails, it is difficult to determine the cause of problem. Hence to successfully explore and implement the deep pipelining and parallelism of our design and use resources in an efficient manner, our testing and validation has been done completely in \textbf{verilog} using \textbf{Vivado} tool.
\subsection{Experimental Setup}
\begin{itemize}
\item \textbf{FPGA :} Our design has been implemented on FPGA board Virtex-7 XC7V690T (on-chip BRAM of 6.46MB, 3600 DSP slices and 693120 logic cells) with a working frequency of 120MHz. This is the same board as used in \cite{Fused layer cnn} and \cite{optimizing fpga}, so that our comparisons in the next section are fair. We have used Xilinx Vivado 2017.1 tool for synthesis, placement and routing and the results are shown in Table (\ref{table:resource}).
\item  \textbf{Baselines:}
We compare our design with the following baselines:
\begin{itemize}
\item \textbf{CPU-caffe:}
We have obtained the baseline CPU-caffe timings with respect to a 3.5GHz hexa-core Intel Xeon E7 caffe-implementation \cite{caffe}.
\item \textbf{GPU-caffe:}
 We have obtained the baseline GPU-caffe timings with respect to GeForce GTX 1070 (1506 MHz graphics clock and 1683MHz processor clock) caffe-implementation \cite{caffe}.
\item \textbf{Fused layer cnn} and \textbf{Optimized convolution accelerator:}
We have compared the resources and timing of first five layers of VGG-16 for DeCoILFNet accelerator against the Fused layer cnn accelerator \cite{optimizing fpga} and Optimizing FPGA-based Accelerator Design for Deep Convolutional Neural Networks \cite{optimizing fpga} by using data from Table of \cite{Fused layer cnn}.
\end{itemize}
\item \textbf{Functional verification:} We performed layer by layer functional verification of our code by comparing it with our Matlab forward pass implementation using trained weights from caffe.  
\end{itemize}

\subsection {Results and Comparison}
In this section, we have analyzed the performance of our accelerators with caffe-cpu, caffe-gpu and  state-of-the-art FPGA-accelerators for the initial layers of VGG-16. The motivation for us to choose VGG-16 was  because modern state-of-the-art deep networks for various applications such as 
Fully Convolutional Network (FCN-32s) \cite{fcn}, Segnet (web demo model) \cite{segnet} are variants of VGG-16. The common feature between them is that most of the convolution layers have kernel size=3X3, padding =1 and stride=1. Also these networks are characterized by multiple consecutive convolution layers.

\begin{table}[!h]
\centering
\caption{Resource Utilization of our accelerator for first 2 convolution layers and 1 pooling of VGG-16}
\begin{adjustbox}{width=\linewidth}
\begin{tabular}{|c|c|c|c|c|}
\hline
Resource & DSP & BRAMs & LUTs & Flipflop  \\
\hline
Used &  605& 474 & 245138 & 465002\\
\hline  
Available & 3600& 1470& 433200 & 866400 \\
\hline
Utilization & 16.8\% & 32.24\% & 56.58\% & 53.67\% \\
\hline
\end{tabular}

\label{table:resource}
\end{adjustbox}
\end{table}

\par
We first evaluate our performance with respect to CPU-caffe and GPU-caffe implementations for the first seven layers of VGG-Net16 ( 5 Convolution Layers and 2 Pooling layers ). Table (\ref{table:VGG_GPU_CPU}) shows the comparison of timing after every layer of VGG-Net16 of our accelerator with software implementations running over both CPU and GPU. As visible from the table, our DeCoILFNet’s performance at 120MHz is comparable to GPU and outperforms CPU with a speedup ranging from 4.28X to 39.08X.\par 
The amount of speedup gained by DeCoILFNet as compared to the CPU keeps on increasing with the increase in the number of layers, this is because of the exploitation of the inter-layer fusion in case of hardware accelerator which allows it to start the next convolution without waiting for the whole output and as the number of layer increases this amount of fusion increases resulting in better performance as compared to the CPU.


\begin{table}[!tbh]
\centering
\caption{Comparing the time taken by first seven layers of VGGNet-16 with CPU-caffe and GPU-caffe. \emph{(Here X is the time taken by DeCoILFNet)}}
\begin{adjustbox}{width=\linewidth}
\begin{tabular}{|c|c|c|c|c|}
\hline
Starting Layer & Ending Layer & CPU-caffe (ms) & GPU-caffe (ms) & DeCoILFNet (ms) \\
\hline
\multirow{2}{*}{conv1\_1}& \multirow{2}{*}{conv1\_1} & 114.54 & 23.12 & 26.76\\ 
\hhline{~~---} && 4.28X & 0.86X &X \\
\hline
\multirow{2}{*}{conv1\_1}& \multirow{2}{*}{conv1\_2} & 736.78 & 27.42 & 27.01 \\
\hhline{~~---}&& 27.27X & 1.01X &X \\
\hline
\multirow{2}{*}{conv1\_1} &\multirow{2}{*}{pool1} & 769.37 &  27.15 & 27.06 \\
\hhline{~~---}&& 28.43X & 1.003X &X \\
\hline
\multirow{2}{*}{conv1\_1} & \multirow{2}{*}{conv2\_1} & 1011.71 & 29.31 & 28.08 \\
\hhline{~~---}&& 36.02X & 1.04X &X \\
\hline
\multirow{2}{*}{conv1\_1} & \multirow{2}{*}{conv2\_2} & 1282.42 & 33.45 & 41.46 \\
\hhline{~~---} && 30.93X & 0.806X &X \\
\hline
\multirow{2}{*}{conv1\_1} & \multirow{2}{*}{pool2} & 1442.47 & 33.57 & 41.49 \\
\hhline{~~---} && 34.76X & 0.809X &X \\
\hline
\multirow{2}{*}{conv1\_1} & \multirow{2}{*}{conv3\_1} & 1637.43 & 34.81 &  41.95\\
\hhline{~~---}&& 39.03X & 0.829X &X \\
\hline
\end{tabular}
\label{table:VGG_GPU_CPU}
\end{adjustbox}
\end{table}

\begin{figure}
\centering
  \includegraphics[width=\linewidth,height = 5cm]{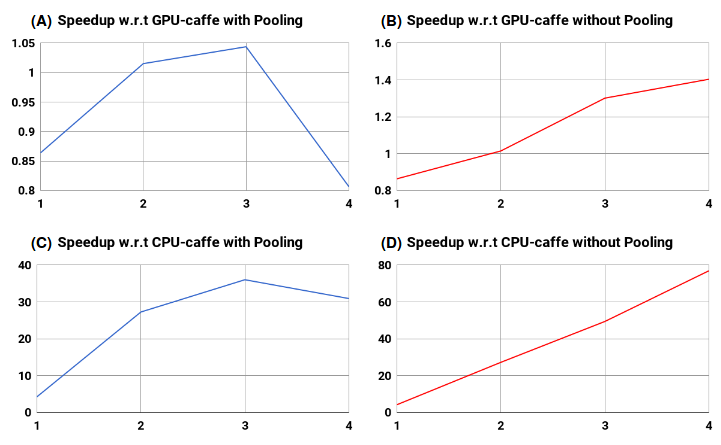}
\caption{Comparison of speedup of our accelerator when compared with GPU-caffe and CPU-caffe with and without pooling layer \textit{(X-axis represents the number of layers, Y-axis represents the speedup)}} \label{fig:pooling}
\end{figure}

Fusing of a pooling layer with convolution layer takes longer than fusing two convolution layers. Fig. (\ref{fig:pooling})  shows the difference in speedup obtained with and without the pooling layer. This is because for computing the pooled layer output, we need to fill up the entire line buffer initially. Thus the initial latency for pooling is higher. \par 
Our design gives the best speedup performance when we have multiple consecutive convolutions. This is particularly helpful in networks like FCNs \cite{fcn} and segnet\cite{segnet} which follow this pattern. In order to demonstrate the performance of our hardware accelerator, we have designed our own network consisting of four consecutive convolution layers each consisting of 64 filters of dimension 3*3 with stride 1, and run it over the CPU, GPU and DeCoILFNet comparing the result after each layer. This is a network pattern that is common in the initial layers of modern networks\cite{segnet}\cite{fcn}.
 As shown in the Table  (\ref{table:consecutive convolution}), when we fuse consecutive convolution layers, we are able to attain a speedup of 76.8X with respect to CPU and even slightly surpasses the GPU speed. In general FPGAs have a much higher per watt performance compared to GPUs. Modern GPUs use 10-100X more power than FPGAs. Thus using a resource constrained FPGA even reaching the GPU computation speed increases the per watt performance significantly.

\begin{table}[!h]
\centering
\caption{Comparing convolution network performance with CPU-caffe and GPU-caffe for consecutive convolution layers.
}
\begin{adjustbox}{width=\linewidth}
\begin{tabular}{|c|c|c|c|c|}
\hline
  Starting Layer & Ending Layer & CPU (ms) & GPU (ms) & DeCoILFNet (ms) \\
\hline
\multirow{2}{*}{Conv\_1} & \multirow{2}{*}{Conv\_1} & 114.54 & 23.12 & 26.764 \\
\hhline{~~---} && 4.28X & 0.863X &X \\
\hline  
\multirow{2}{*}{Conv\_1} & \multirow{2}{*}{Conv\_2} & 736.78 & 27.42 & 27.01 \\
\hhline{~~---} && 27.27X & 1.015X &X \\
\hline
\multirow{2}{*}{Conv1\_1} & \multirow{2}{*}{Conv1\_3} & 1346.32 & 35.45 & 27.24\\
\hhline{~~---} && 49.42X & 1.301X &X \\
\hline
\multirow{2}{*}{Conv1\_1} & \multirow{2}{*}{Conv1\_4} & 2113.24 & 38.58 & 27.48 \\
\hhline{~~---} && 76.91X & 1.403X &X \\
\hline
\end{tabular}

\label{table:consecutive convolution}
\end{adjustbox}
\end{table}

In order to compare our architecture with the current state-of-the-art hardware accelerators we compared our architecture with the one proposed by \cite{optimizing fpga} and \cite{Fused layer cnn} for the first seven layers of VGG-Net16. Table \ref{table:accelerator} compares the resource utilization of DeCoILFNet with the baseline architectures. The resource utilization and timing for both implementations has been taken directly from \cite{Fused layer cnn}). Among the three, our architecture gives the best performance of speed(compared to \cite{optimizing fpga}\cite{Fused layer cnn}) along with a significant reduction in data volume transferred (compared to \cite{optimizing fpga}. We have been able to effectively utilize the DSPs by eliminating recomputation with the help of line-buffer pipelining. The goal of our architecture is to maximize the speedup in limited external memory accesses. Depth concatenation helped us pipeline dataflow and perform all independent computations for the first seven layers of VGG \cite{VGGNet} in parallel. The pipelining is also very stringent, i.e. there is no stall after the initial latency and we keep getting a continuous stream of output. Keeping these in mind, the results shown in Table\ref{table:accelerator} are the best possible we could achieve on Virtex-7. We are able to attain more than 2X speedup in terms of clock cycles compared to both accelerators, along with higher working frequency.

\begin{table}[!h]
\centering
\caption{Comparison table with FPGA accelerators for initial layers of VGG-Net:}
\begin{adjustbox}{width=\linewidth}
\begin{tabular}{|c|c|c|c|}
\hline
 & Optimized & Fused Layer & DeCoILFNet \\
\hline
Clockcycles*$10^3$ & 10951 & 11655 & 5034 \\
\hline
Precision & 32 bits float & 32 bits float & 32 bits fixed \\
\hline  
Frequency(in MHz) & $100$ & $100$ & $120$ \\
\hline
MB transferred per input & 77.14 & 3.64 & 6.69 \\
\hline
BRAMs & 2085 & 2509 & 2387 \\
\hline
DSP & 2880 & 2987 & 2907 \\
\hline
\end{tabular}

\label{table:accelerator}
\end{adjustbox}
\end{table}

\section{Discussion and Trade-off}
\begin{figure}[!tbh]
\centering
  \includegraphics[width=\linewidth,height = 4 cm]{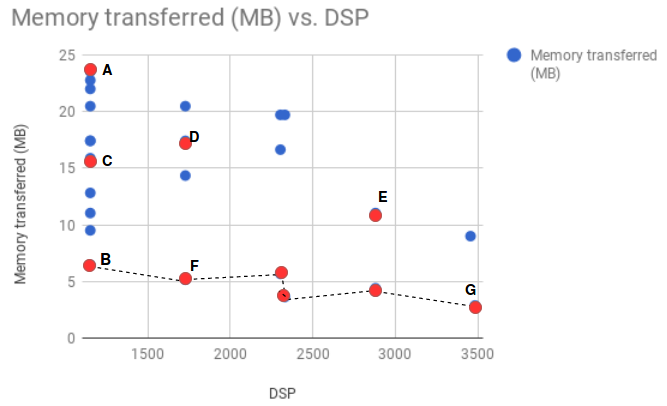}
\caption{Trade-off between inter-layer fusion and computation resource: DSP}\label{fig:Tradeoff}
\end{figure}

Fig. (\ref{fig:Tradeoff}) shows the relation between off-chip memory accesses and the computation units due to grouped fusion of five convolutions and two pooling layers of VGG-16 in different groups. We have assumed that the depth based parallelism is constant for all the cases considered. The point A represents no fusion, i.e. when all intermediate outputs are stored in DDR. In this case as is visible from the diagram, since we write back to the DDR, the computation unit of single layer is reused for every layer, i.e. each layer is its own group. Hence the DSP utilization is minimum for this case at the cost of highest ( 23.54 MB ) dataflow. The point G in the diagram represents when all layers have been grouped and fused. Since we are computing all layers concurrently, the DSP utilization is maximum with minimum on-chip memory utilization. 


\par Our high performance in Table \ref{table:accelerator} compared to other accelerators is aided by our parallel computations across depth. Using depth concatenation allows us to perform several computations concurrently. Our depth-concatenation technique is also limited by the compute resources present on the FPGA board. As the concatenation depth increases, we need more resources to perform computations in parallel. We have used iterative decomposition to solve this problem. We divide the depth into multiple groups of parallel computation, and process these groups serially. The number of serial groups decides the factor by which our clock cycles increase, as we need to wait for the result of all groups to complete to get one output. This technique is particularly needed for the later layers of VGG-Net where we need to process inputs of depth 256 or 512.  
\par
In CNNs, the input and output feature volume is large for initial layers and gradually reduces. Keeping data-volume considerations aside, independence in computation-pattern for later-layers is same as initial-layers. Though we have demonstrated improvement results for initial layers, we believe our architecture can exploit the same data independence of later layers to give same better performance over baselines. For later-layers, weights dominate memory space and depth of convolving filters increases significantly. Since both parallelization due to depth concatenation and layer fusion require same compute resources, there is a trade-off between them. The number of layers fused should be maximum for the initial layers. This is because for the initial layers, the intermediate output data is huge and less layers fused would mean a huge data volume movement to and from external memory \cite{Fused layer cnn}. Whereas for the later layers, the depth of input and convolving filters increases significantly and the intermediate. Also the subsampling layers reduce the intermediate data volume. Hence it makes more sense to allocate compute resources to parallel computations across depth for the later layers.  
  
\section{Conclusion}
We presented a ‘Depth Concatenation and inter-layer fusion based convnet accelerator-DeCoILFNet’ which exploits the intra-layer parallelism of CNNs by flattening across the depth and combining it with the inter-layer fusion.  Our accelerator maximises data re-use and completely eliminates recomputations while fusing multiple convnet layers. We explained in detail the different components of our architecture and evaluated our accelerator on VGG-like networks, with VGG-16 as the representative. We demonstrated that our 120 MHz accelerator is 30X faster compared the performance to a 3.5GHz hexa-core Intel Xeon E7 caffe-implementation.In addition, our design reduces external memory access by 42X along with a speedup of more than 2X in the number of clock cycles compared to state-of-the art FPGA accelerators.

\end{document}